\def\year{2021}\relax
\documentclass[letterpaper]{article} 
\usepackage{aaai21}  
\usepackage{times}  
\usepackage{helvet} 
\usepackage{courier}  
\usepackage[hyphens]{url}  
\usepackage{graphicx} 
\usepackage{natbib}  
\usepackage{caption} 
\usepackage{nameref}
\usepackage{cite}
\usepackage{amsmath,amssymb,amsfonts}
\usepackage{algorithmic}
\usepackage{graphicx}
\usepackage{textcomp}
\usepackage[hyphens]{url}

\usepackage{subcaption}

\usepackage{xcolor}
\def\BibTeX{{\rm B\kern-.05em{\sc i\kern-.025em b}\kern-.08em
    T\kern-.1667em\lower.7ex\hbox{E}\kern-.125emX}}
\begin{document}

\title{Privacy in Crisis: A Study of Self-Disclosure During the Coronavirus Pandemic}

\author{
Taylor Blose, Prasanna Umar, Anna Squicciarini, Sarah Rajtmajer \\
}

\affiliations{
College of Information Sciences and Technology, The Pennsylvania State University, USA \\
thb5018@psu.edu, pxu3@psu.edu, acs20@psu.edu, smr48@psu.edu
}

\maketitle

\begin{abstract}
We study observed incidence of self-disclosure in a large dataset of Tweets representing user-led English-language conversation about the Coronavirus pandemic. Using an unsupervised approach to detect voluntary disclosure of personal information, we provide early evidence that situational factors surrounding the Coronavirus pandemic may impact individuals' privacy calculus. Text analyses reveal topical shift toward supportiveness and support-seeking in self-disclosing conversation on Twitter. We run a comparable analysis of Tweets from Hurricane Harvey to provide context for observed effects and suggest opportunities for further study.
\end{abstract}

\section{Introduction}

\noindent At the time of writing, one-third of the world's population is directly impacted by restrictions on movement and activity, aimed at slowing the Coronavirus pandemic. 
All fifty states of the U.S. have issued guidelines on permissible travel outside the home, limitations on public gatherings, and business closures. 
Experts suggest that some of these measures may continue for 18 months or more. 
 
Unsurprisingly, there has been an unprecedented surge in online activity \cite{GovTech}. 
Much of the increased traffic extends beyond typical Internet surfing and video streaming, as people find ways to leverage online resources to stay connected with one another, personally and professionally. 
Video-conferencing and social media usage are soaring as people turn to these platforms to support routine social activities \cite{ZoomUsers,FacebookLimits}.

It is to be expected that the expanded breadth and depth of online activity  will magnify privacy risks for individual users.
In addition to simply spending more time online, the emergence of virtual play dates and book clubs suggests that during this time of social distancing, people are looking for ways to stay close (e.g., "apart but together"). This may be particularly the case for the many individuals facing heightened anxiety, stress and depression due to social isolation, grief, financial insecurity, and of course health-related fears of the virus itself 
\cite{APA, CNBC}.

The literature on social communication suggests that interpersonal connectedness and relationship development is fundamentally facilitated through iterative \emph{self-disclosure} \cite{altman1973social}, that is, intentionally revealing personal information such as personal motives, desires, feelings, thoughts, and experiences to others \cite{derlaga1987self}. In fact, there is a robust literature on (routine) self-disclosure in online social media outside the particular domain of crisis \cite{bak-etal-2014-self-disclosure,joinson2007self,nguyen2012comparing,houghton2012linguistic,Attrill2011}. 
Research indicates that as users engage in discussion online, they leverage self-disclosure as a way to enhance immediate social rewards \cite{HZ17}, increase legitimacy and likeability \cite{BKO12}, and derive social support \cite{TW02}. 
Despite the ``upsides", i.e., socially adaptive motivations for disclosure, we know that self-privacy violations can come at a cost, leaving users exposed to 
identity theft, cyber fraud and other crimes \cite{hasan2013discussion}, discrimination in job searches, credit and visa applications \cite{Conversation}, harassment and bullying \cite{peluchette2015cyberbullying}. Indeed, studies have repeatedly shown that an overwhelming majority of users have privacy concerns about their online interactions \cite{smith2011information}. These concerns evolve over time, tied to the day’s events and the longer arc of shifting norms \cite{acquisti2015privacy, adjerid2016beyond}. 

Little is known about the evolution of users' sharing practices during crisis. We know that victims of Hurricane Harvey sought assistance through social media, in some cases revealing their full names and addresses online \cite{WSJ}. However, what we are witnessing in the case of the Coronavirus pandemic is distinct from previous crises in important ways. COVID-19 is a global, relatively protracted acute threat. Unlike natural disasters or military engagements, the pandemic has left communication infrastructure intact. Digital outlets have become lifelines. We posit that self-privacy violations during the Coronavirus pandemic can help individuals feel more socially connected during a time of anxiety and physical distance, even though the long term impact of these disclosures is unknown. 
 
In this work, we carry out analysis of instances of self-disclosure in a dataset of 53,557,975 Tweets representing conversations on Coronavirus-related topics. We leverage an unsupervised method for identifying and labeling voluntarily disclosed personal information, both subjective and objective in nature. 
Our main finding reveals a steep increase in instances of self-disclosure, particularly related to users' emotional state and personal experience of the crisis. To our knowledge, this work is the first to study self-disclosure on social media during the Coronavirus pandemic. We run comparative analyses on a dataset of Tweets representing conversations about Hurricane Harvey in late summer 2017. The Harvey study provides valuable context for these unprecedented times, suggesting similarities and differences that better inform our observations during the current crisis and the privacy and crisis literatures in general.

\section{Related work}


Research in the space of online interactions has sought to understand the actualization of self-disclosure in digitally-mediated social communication. Studies suggest that disclosure behaviors in online environments may be meaningfully different than their offline counterparts, e.g., anonymity and lack of nonverbal cues afforded by social media may encourage greater disclosure of sensitive information \cite{forest2012social, joinson2001self}. Similar findings are reported in \cite{Ma2016}, where authors explore the impact of content intimacy on self-disclosure. It is well-established for face-to-face communication that people disclose less as content intimacy increases, but this effect seems to be weakened in online interactions.

Recent work has positioned online self-disclosure as strategic behavior targeting social connectedness, self-expression, relationship development, identity clarification and social control \cite{abramova2017understanding,bazarova2014self,de2014mental}. Voluntary disclosure of personal information has been associated with improved well-being, meaningfully related to increased informational and emotional support \cite{Huang2016}. 

There are, however, potential costs to users' privacy, including the unforeseen use and sharing of disclosed data. Early work by Acquisti and Gross \cite{acquisti2006imagined} suggested that social network users were neither fully aware nor responsive to privacy risks. Over time, studies have captured a shift toward increased privacy awareness \cite{johnson2012facebook,Vitak2014}, but there remains great variability in information sharing behaviors amongst individuals and across platforms  \cite{zhao2016social}. It has been shown that culture plays a role in disclosure decisions \cite{ZhaoSD, krasnova2012self,trepte2017cross}, as does gender \cite{Sun2015} and socioeconomic status (SES) \cite{marwick2017nobody}. 
Overarchingly, the cost-benefit analyses underlying an individual's decision to share in the presence of privacy risk is postulated by social exchange theory \cite{emerson1976social} and re-framed in the context of online social networks as the so-called privacy calculus \cite{krasnova2010online,dienlin2016extended}.

Critically, work in a number of domains suggests that contextual and situational factors, e.g., trust, anonymity, financial incentives, are embedded within the privacy calculus \cite{Joinson2012,hann2007overcoming,li2010understanding}. Amongst these factors, emotion has also been suggested to play a meaningful role in privacy behaviors \cite{laufer1977privacy,berendt2005privacy,li2017resolving}. This finding is in keeping with the general theory of feeling-as-information \cite{petty2001role}, whereby emotions serve as information cues directly invoking adaptive behaviors \cite{lazarus1991emotion}.
Studies linking emotion to disclosure online have thus far limited scope to considering the emotional impact of a particular website. The ways in which an individual's mood and general emotional state impact individual privacy calculus are unknown.

To our knowledge, there is no literature examining changes to patterns of self-disclosure in crisis through the lens of privacy risk. The crisis community is interested in the related but fundamentally distinct problem of mining self-disclosed information for the purposes of identifying and deploying assistance and relief to impacted individuals and communities (see \cite{muniz2020} for review).

This work also dovetails with the literature on detection and tagging of self-disclosure in text, e.g., \cite{CaliskanIslam:2014:PDD:2665943.2665958,Wang:2016:MSS:2818048.2820010,vasouludoi:10.1002/asi.21610,bak-etal-2014-self-disclosure,Chow:2008:DPL:1401890.1401997,Choi:jucs_19_16:text_analysis_for_monitoring}. Chow et. al. \cite{Chow:2008:DPL:1401890.1401997} developed an association rules-based inference model that identified sensitive keywords which could be used to infer a private topic. Similarly, multiple studies utilized pattern or rule based methods to detect specific types of disclosures \cite{vasouludoi:10.1002/asi.21610,Umar2019}. 

Past work has attempted to classify self-disclosure by levels, or degree of disclosure. Caliskan et.al \cite{CaliskanIslam:2014:PDD:2665943.2665958} used AdaBoost with Naive Bayes classifier to detect privacy scores for Twitter users' timelines.  Bak et.al. \cite{bak-etal-2014-self-disclosure} applied modified Latent Dirichlet Allocation (LDA) topic models for semi-supervised classification of Twitter conversations into three self-disclosure levels: general, medium and high. Wang et. al. \cite{Wang:2016:MSS:2818048.2820010} used regression models with extensive feature sets to detect degree of self-disclosure. Because the notion of sensitive information is based on user perception and context, studies on detection of self-disclosure levels are often difficult to generalize beyond their original context.

\section{Primary dataset}

Our primary dataset is a repository of Tweet IDs corresponding to content posted on Twitter related to the Coronavirus pandemic \cite{CVD19_Twitter}. At the time of writing, the repository contains  508,088,777 Tweet IDs for the period of activity from January 21, 2020 through August 28, 2020. Tweets were compiled utilizing a combination of Twitter's Search API (for activity January 21 through January 28) and Streaming API (for activity January 28 through July 31). The repository represents topically-relevant Tweets across the platform, canvassed based on designated keywords, as well as the full activity of selected accounts (See Tables \ref{keywords}, \ref{accounts}). Around June 6th, the repository collection infrastructure transitioned to Amazon Web Services which generated a significant increase in Tweet ID volume. No search parameters were adjusted or data gaps presented because of the transition; therefore, we analyzed the entirety of the dataset in a consistent manner. In analyses that follow, we focused our scope to highlight an important transitional period in the pandemic for most users in the United States by considering two sub-periods -- prior to and after March 11th, the date of the World Health Organization's pandemic declaration and just two days preceding U.S. President Trump's declared national emergency.

Text and metadata corresponding to these 508,088,777 Tweet IDs were obtained through rehydration using the Twarc\footnote{https://github.com/DocNow/twarc} Python library. Of the IDs passed for rehydration,  461,259,923 were successfully rehydrated. The 9.21\% loss represents deleted content, therefore irretrievable through Twitter's API.

For the purpose of measuring and studying self-disclosure, we filtered the corpus to capture Tweets that represent original content posted by individual users. Specifically, we removed quoted Tweets, retweets,\footnote{Retweets were identified through the existence of the ``retweetedstatus'' field in the Tweet object returned by the API. Tweets beginning with the string 'RT @' were also treated as retweeted records.} as well as all Tweets associated with verified accounts and the specific organizational accounts listed in Table \ref{accounts}. We narrowed our analysis to English-language content in order to reduce situational heterogenity and maintain confidence in our labeling approach, which has been developed and validated on English-language text. The resulting corpus, which forms the basis of our analyses, consists of 53,557,975 unique Tweets.


\begin{table}[htbp]
\centering
\begin{tabular}{|p{6.5cm}|p{1.5cm}|}
\hline
\textbf{Keyword Followed}&\textbf{Start Date} \\
\hline
Coronavirus, Koronavirus, Corona, CDC,\newline
Wuhancoronavirus, Wuhanlockdown, Ncov, \newline
Wuhan, N95, Kungflu, Epidemic, outbreak, \newline
Sinophobia, China & 1/28/2020 \\
\hline 
covid-19 & 2/16/2020\\
\hline 
corona virus & 3/2/2020\\
\hline
covid, covid19, sars-cov-2 & 3/6/2020\\
\hline
COVID-19 & 3/8/2020\\
\hline
COVD, pandemic & 3/12/2020\\
\hline
coronapocalypse, canceleverything, \newline
Coronials, SocialDistancingNow, 
\newline Social Distancing, 
SocialDistancing & 3/13/2020\\
\hline
panicbuy, panic buy, panicbuying, \newline
panic buying, 14DayQuarantine, \newline
DuringMy14DayQuarantine, panic shop, \newline
panic shopping, panicshop, \newline
InMyQuarantineSurvivalKit, panic-buy, \newline
panic-shop & 3/14/2020\\
\hline 
coronakindness & 3/15/2020\\
\hline
quarantinelife, chinese virus, chinesevirus, \newline
stayhomechallenge, stay home challenge, \newline
sflockdown, DontBeASpreader, lockdown, \newline
lock down & 3/16/2020\\
\hline
shelteringinplace, sheltering in place, \newline
staysafestayhome, stay safe stay home, \newline
trumppandemic, trump pandemic, \newline
flattenthecurve, flatten the curve, \newline
china virus, chinavirus & 3/18/2020\\
\hline
quarentinelife, PPEshortage, saferathome, \newline
stayathome, stay at home, stay home, \newline
stayhome & 3/19/2020\\
\hline
GetMePPE & 3/21/2020\\
\hline
covidiot & 3/26/2020\\
\hline
epitwitter & 3/28/2020\\
\hline
pandemie & 3/31/2020\\
\hline 
wear a mask, wearamas, kung flu, covididiot & 6/28/2020\\
\hline
COVID\_\_19 & 7/9/2020 \\
\hline
\end{tabular}
\caption{Keywords followed, by start date}
\label{keywords}
\end{table}

\begin{table}[htbp]
\centering
\begin{tabular}{|p{6.5cm}|p{1.5cm}|}
\hline
\textbf{Account Followed}&\textbf{Start Date} \\
\hline
PneumoniaWuhan, CoronaVirusInfo,V2019N, \newline
CDCemergency, CDCgov, WHO, HHSGov, \newline
NIAIDNews  & 1/28/2020 \\
\hline 
drtedros & 3/15/2020 \\
\hline
\end{tabular}
\caption{Accounts followed, by start date}
\label{accounts}
\end{table}

\section{Automated detection of self-disclosure} \label{detection}

\begin{figure*}[tbh]
\centering
\begin{subfigure}{.4\textwidth}
  \centering
  \includegraphics[width=.8\linewidth]{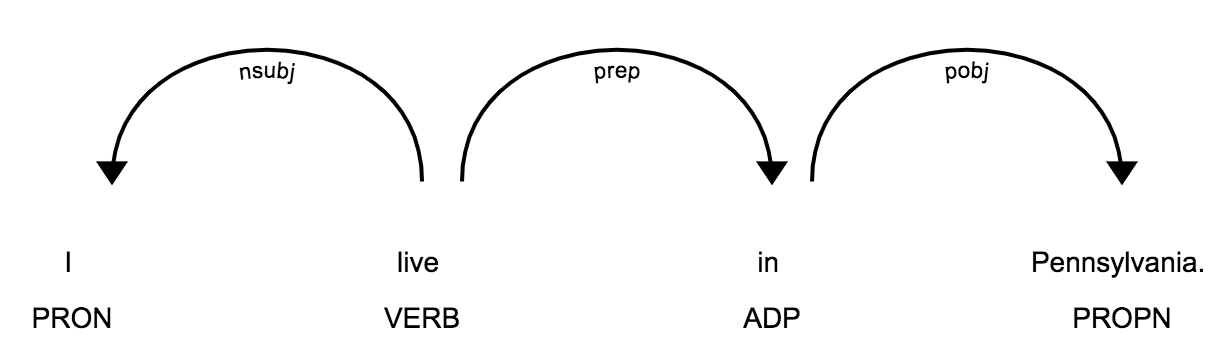}
  \caption{Dependency tree}
  \label{fig:deptree}
\end{subfigure}%
\begin{subfigure}{.4\textwidth}
  \centering
  \includegraphics[width=.8\linewidth]{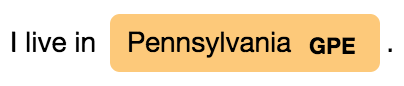}
  \caption{Named entity recognition}
  \label{fig:ner}
\end{subfigure}
\caption{Illustration of phases in self-disclosure categorization scheme \cite{Umar2019}}
\label{example}
\end{figure*}

We use an unsupervised method \cite{Umar2019} to detect instances of self-disclosure in our dataset. Consistent with the literature on detection of self-disclosure in text (see, e.g., \cite{bak-etal-2014-self-disclosure,de2014mental,houghton2012linguistic,Wang:2016:MSS:2818048.2820010}), we consider the presence of first-person pronouns. Specifically, we consider sentences containing self-reference as the subject, a category-related verb and associated named entity. Consider the example of location self-disclosure shown in Figure \ref{example}. A first person pronoun ``I'' is the self-referent subject of the sentence. It is used with a location-related verb ``live'' in the vicinity of the associated location entity ``Pennsylvania''. Notably, subjective categories of self-disclosure such as interests and feelings do not have associated named entities. These are differentiated through rule-based schemas based on subject-verb pairs. The approach described is implemented in three phases: 1) subject, verb and object triplet extraction with awareness to voice (active or passive) in the sentence; 2) named entity recognition; 3) rule-based matching to established dictionaries. Our dictionaries are adopted from \cite{Umar2019}.

As the proposed approach is based on sentence structure and syntactic resources (subject, verb, object and entities), it can be applied to any textual content. However, we acknowledge that Tweets present unique characteristics. Due to character limits and consequent emerging norms of the platform, users more frequently engage acronyms and abbreviations \cite{han2011lexical}. Relatedly, sentence structure and syntax are noisier when compared to more verbose platforms \cite{boot2019character}. 
User mentions, hashtags, and graphic symbols are embedded within text. 
Considering these differences, we pre-processed Tweets as follows. All Unicode encoding errors were corrected. We removed markers associated with retweets (e.g., ``RT'') and filtered user mentions and hashtags. Additionally, symbols like ``\&'' and ``\$'' were replaced with their respective word representations. Email addresses and phone numbers were replaced with placeholders ``emailid'' and ``phonenumber'', while URLs were filtered. We also replaced contractions in the Tweets like ``I'm'' to ``I am'' and corrected incorrect use of spacing between words. These pre-processing steps enabled cleaner input to the detection algorithm. 

We validated our approach on a baseline dataset of 3,708 annotated Tweets \cite{CaliskanIslam:2014:PDD:2665943.2665958}. The dataset was manually labelled for presence or absence of self-disclosed personal information~
such as location, medical information, demographic information and so on. 
For our purposes, category-specific labels of self-disclosure were binarized, resulting in a total of 3,188 self-disclosing and 520 non-self-disclosing Tweets. Using the unsupervised method described, we classified the baseline data and compared against the binarized manual labels. Our approach demonstrated acceptable performance yielding precision, recall and F scores of 92.5\%, 50.3\% and 65.1\% respectively. Recall is lower than reported in \cite{Umar2019}, attributable to differences between the taxonomy used to manually label the baseline Twitter dataset and the taxonomy used to create the unsupervised detection method (e.g., our unsupervised approach was not developed to detect categories like drug use and personal attacks, whereas these categories were explicitly offered to manual labelers). 

\section{Topic modeling} \label{topics}

Latent Dirichlet Allocation (LDA) \cite{blei2003latent} was used for topic modeling of Tweets in our dataset. 
In this approach, each document (a single Tweet) in the corpus (set of Tweets) is considered to be generated as a mixture of latent topics and each topic is a distribution over words. For a document, each word is assigned topics according to a Dirichlet distribution. Iteratively cycling through each word in each document and all documents in the corpus, topic assignments are updated based on the prevalence of words across topics and the prevalence of topics in the document. Based on this process, final topic distributions for documents and word distributions for topics are generated.  

In addition to the cleaning steps described in Section \ref{detection}, we pre-processed Tweets using tokenization, conversion to lower case, lemmatization and removal of punctuation and stopwords. Topics were generated from the resulting corpora using the LDA model within the Gensim Python library.\footnote{https://radimrehurek.com/gensim/models/ldamodel.html}

We ran topic analyses over subsets of interest within the complete Coronavirus dataset. Namely, we explored unique topic models for Tweets in two time windows pre- and post-March 11 (January 21 - March 11; March 12 - May 15), and compared topical themes between Tweets with presence or absence of detected instances of self-disclosure over the entirety of this subset. To better understand disclosure trends presented post-March 11, we also generated topic models over one month windows for the remainder of the Coronavirus dataset: May 16 through June 15, June 16 through July 15, and July 16 through August 15. 



\section{Findings} \label{findings}

\begin{figure}
\centering
\includegraphics[width=0.99\columnwidth]{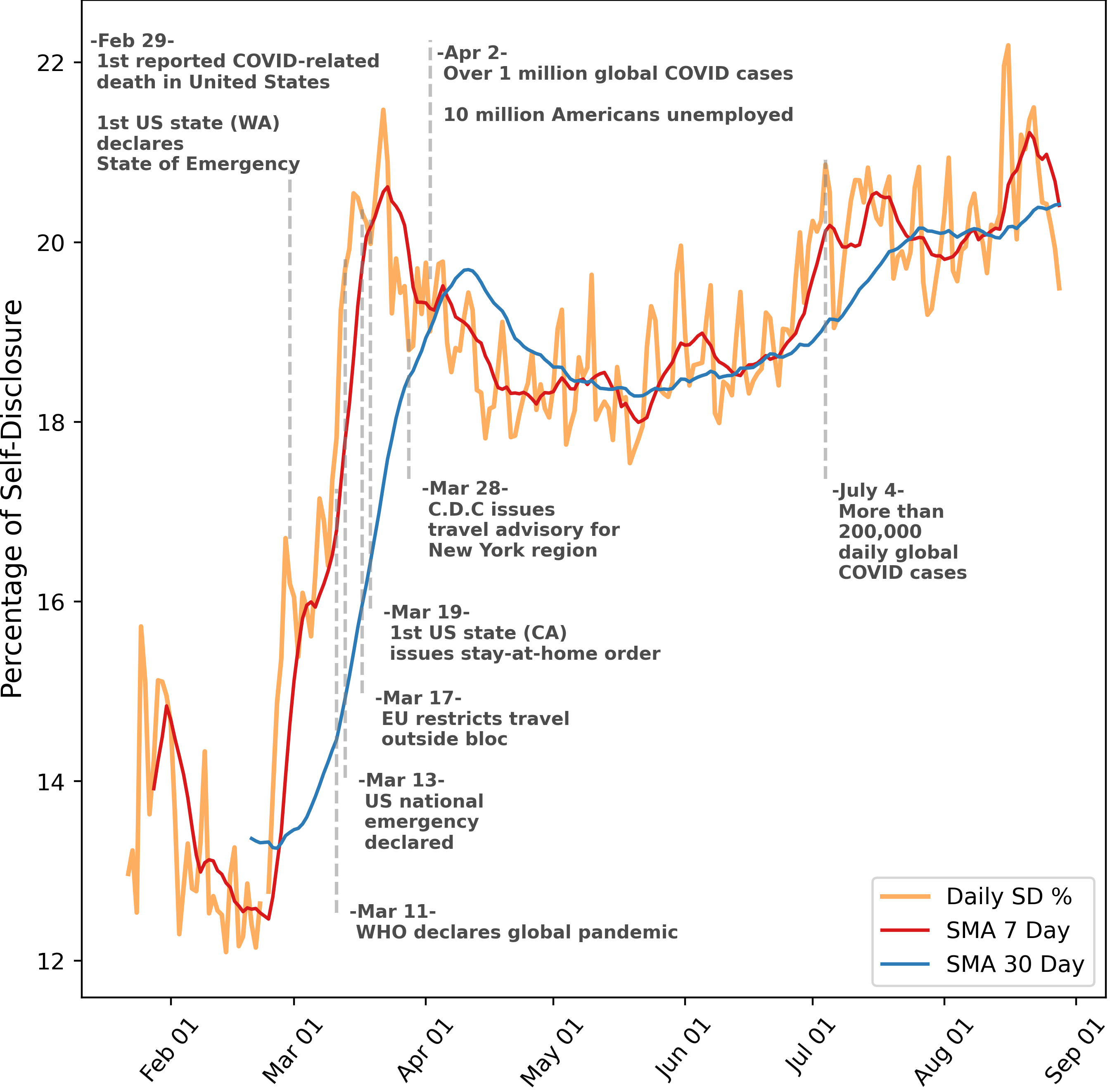}
\caption{Percentage of Tweets containing self-disclosure, assessed daily with 7- and 30-day simple moving averages}
\label{fig:dailySD}
\end{figure}

Of the 53,557,975 Coronavirus-related Tweets analyzed, approximately 19.07\% (10,215,752 Tweets) contain elements of self-disclosure. Looking more closely at daily variance from January until June, we identified a significant transition point in activity around March 11, 2020, as shown in Figure \ref{fig:dailySD}. The significance of the behavioral change is supported by overlaying 7-day and 30-day simple moving averages which smooth day-to-day variance by taking the average disclosure percentage value over the given time window. 

In the period January 21 through March 11, the average daily percentage of self-disclosing Tweets is 14.63\%; from March 12 through May 15, the average daily percentage is 18.89\%. 
This change in activity coincides with an escalation in severity and increased global awareness of the crisis, with the World Health Organization (WHO) officially classifying Coronavirus as a pandemic (March 11). Current events coincident with observed changes in the rate of self-disclosure are noted on March 13 and March 19 when the United States officially declared a national state of emergency, and when the governor of California issued the first statewide `stay-at-home' order, respectively. Self-disclosure activity remains high for the remainder of the dataset with an average daily percentage of 19.79\% from May 16 through August 28. 
These observations suggest that situational context, in particular during crisis, may meaningfully influence short- and long-term disclosure behaviors.


\begin{figure}[tbh]
  \centering
  \begin{minipage}[b]{.9\linewidth}
    \subcaptionbox{January 21 - March 11, 2020 \label{sd_PreMar11}}
      {\includegraphics[width=.975\linewidth]{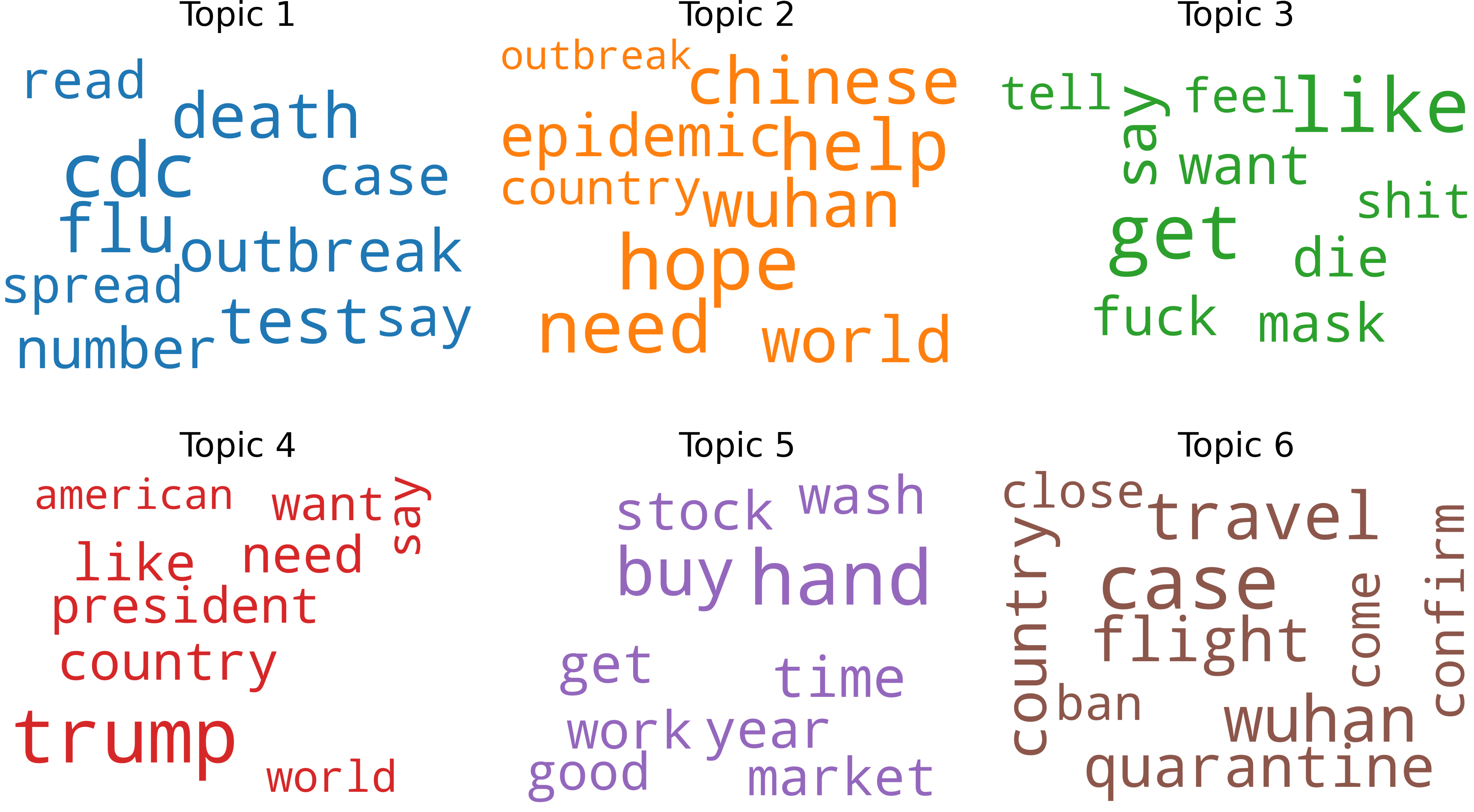}}\\
     
    \subcaptionbox{March 12 - May 15, 2020 \label{sd_PostMar11}}
      {\includegraphics[width=.975\linewidth]{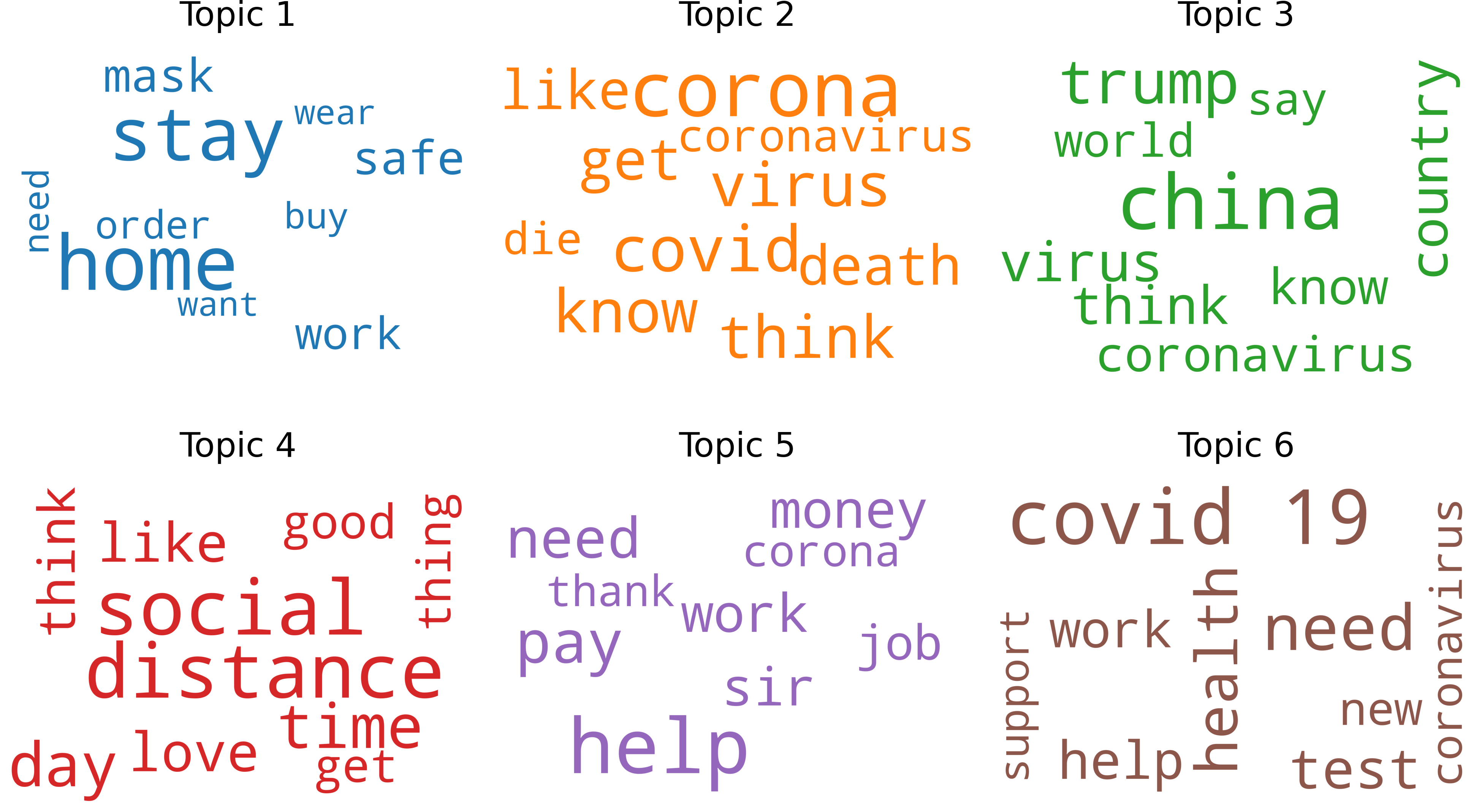}}%
  \end{minipage}%
  \caption
    {%
      Topical Comparison of Self-Disclosing Coronavirus Tweets Pre- \& Post-March 11, 2020%
      \label{COVIDSD_prePost}%
    }%
\end{figure}


\begin{figure*}
\centering
\begin{subfigure}{\columnwidth}
  \centering
  \includegraphics[width=.85\linewidth]{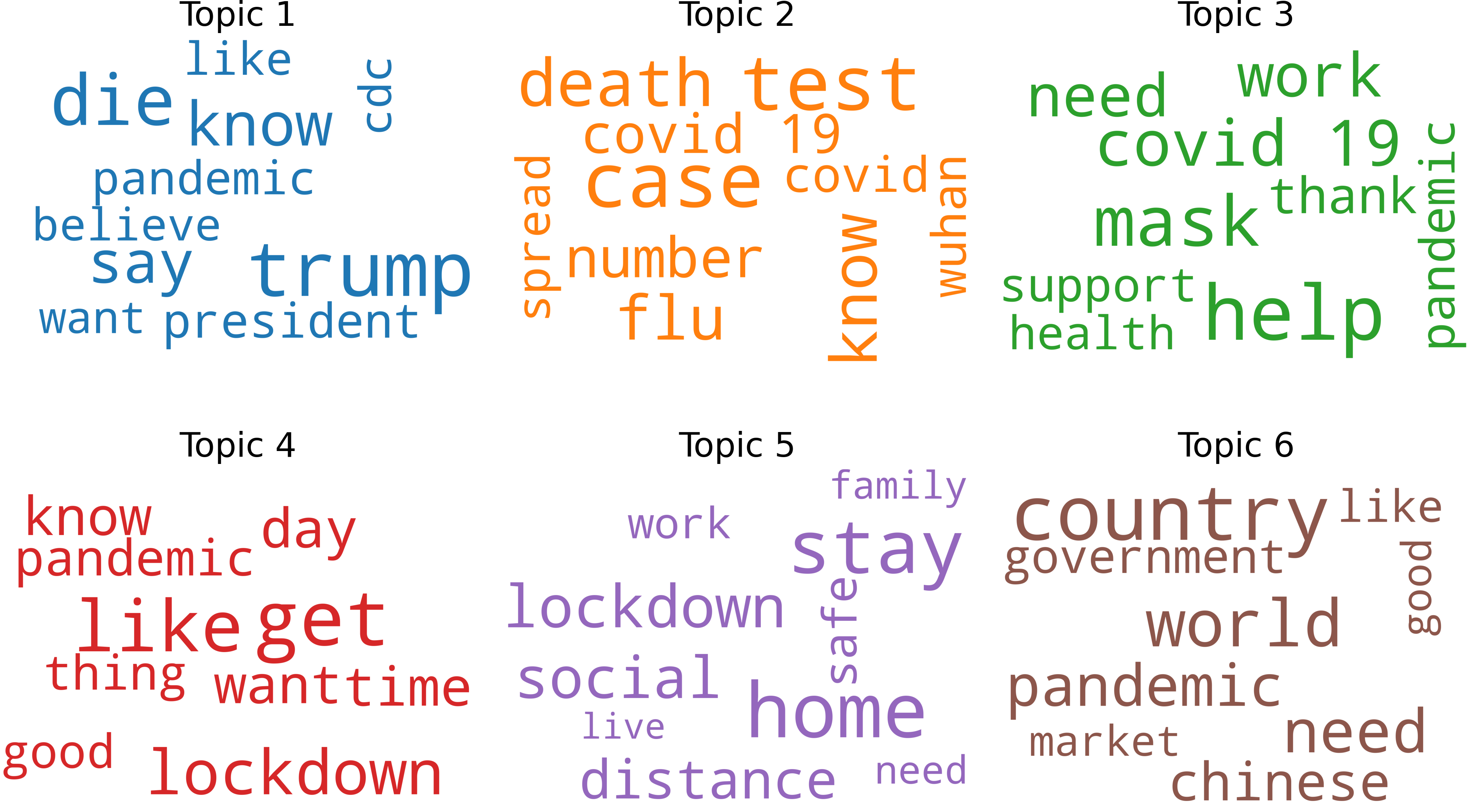}
  \caption{Self-disclosing Tweets}
  \label{fig:sub1_preMar11}
\end{subfigure}
\begin{subfigure}{\columnwidth}
  \centering
  \includegraphics[width=.85\linewidth]{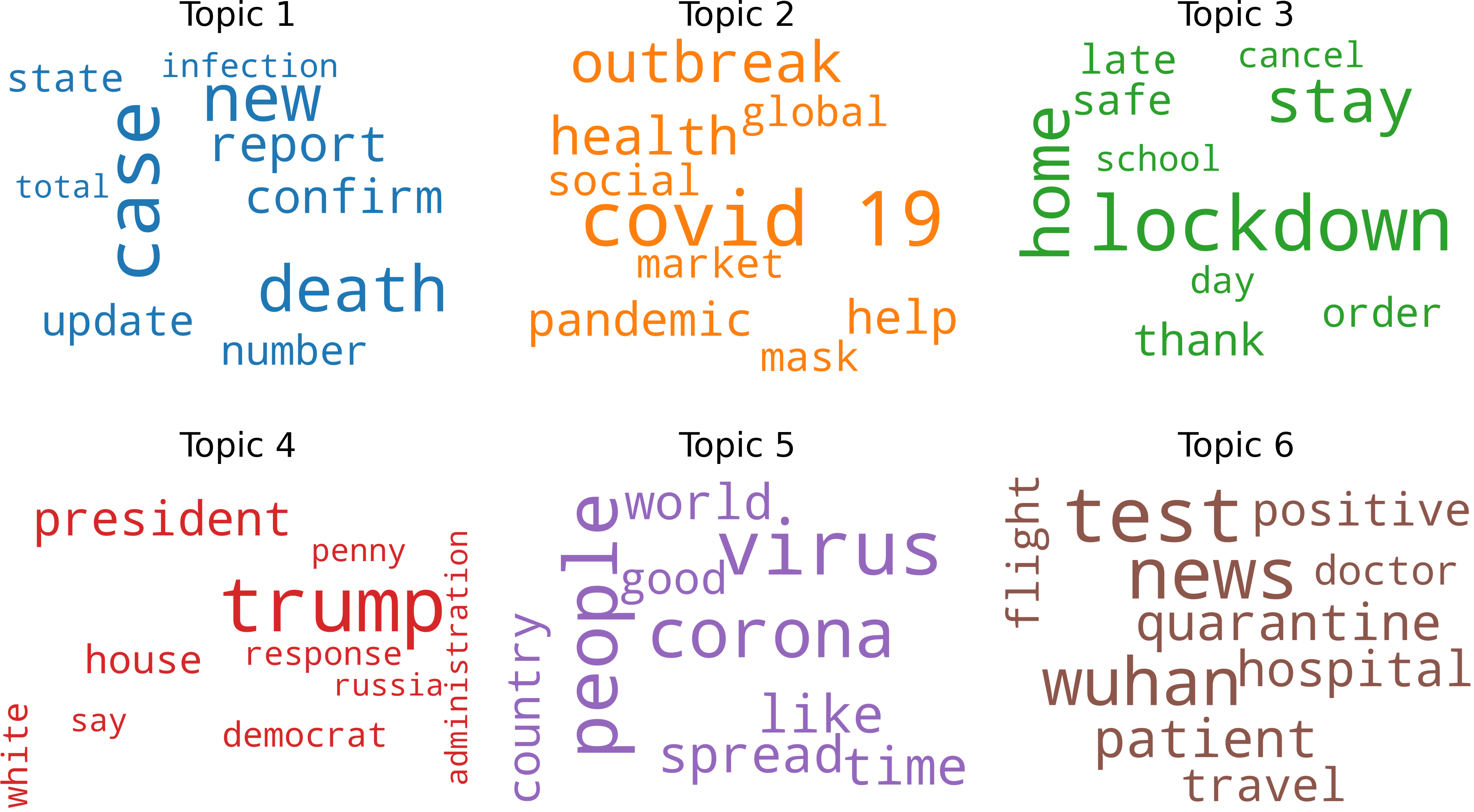}
  \caption{Non-self-disclosing Tweets}
  \label{fig:sub2_postMar11}
\end{subfigure}
\caption{Topical Comparison of Coronavirus Tweets through May 15, 2020}
\label{COVIDSD_subSD_nonSD}
\end{figure*}

\begin{figure*}[tbh]
\centering
\begin{subfigure}{.33\textwidth}
  \centering
  \includegraphics[width=\linewidth]{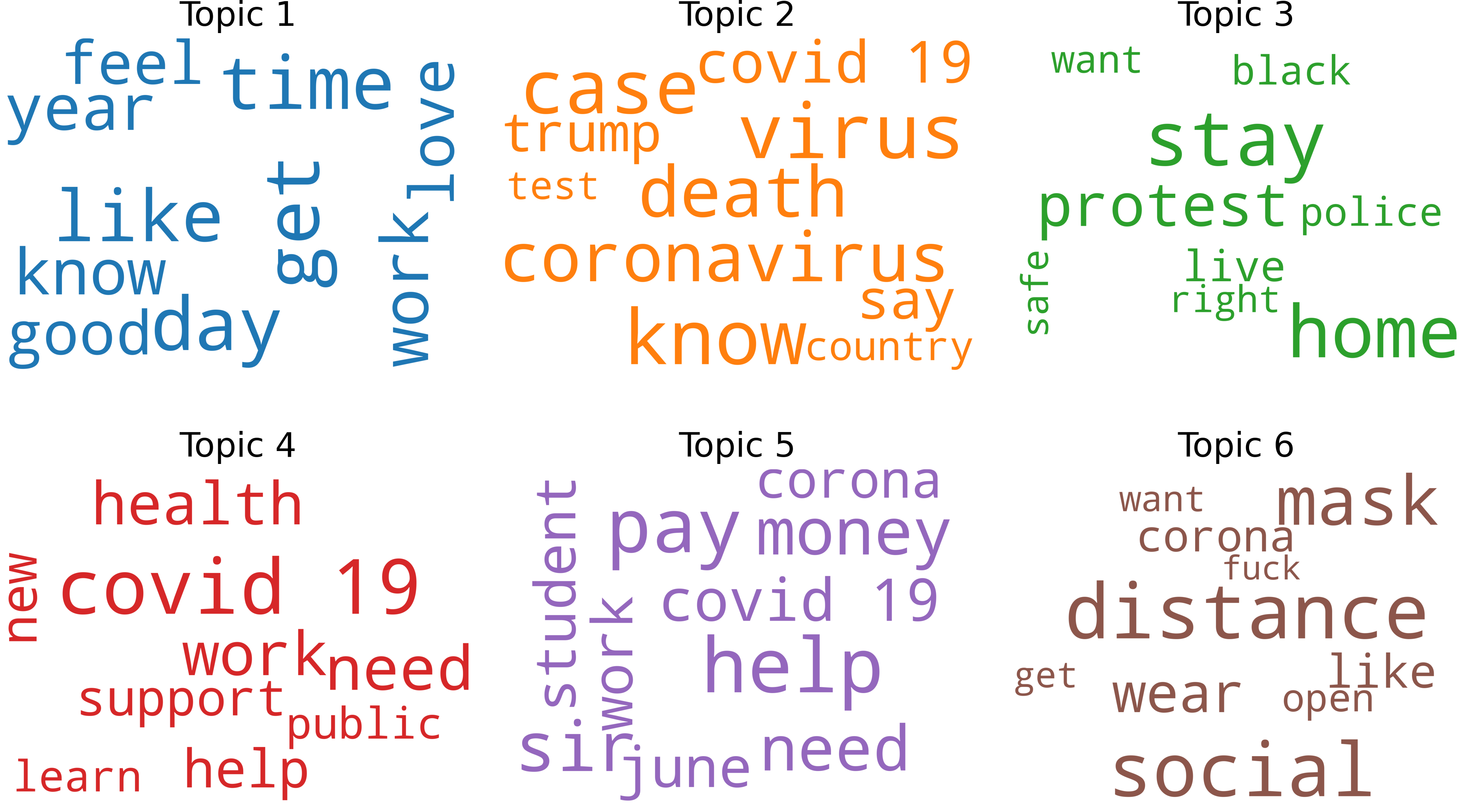}
  \caption{May 16 - June 15, 2020}
  \label{fig:sdMayJune}
\end{subfigure}
\begin{subfigure}{.33\textwidth}
  \centering
  \includegraphics[width=\linewidth]{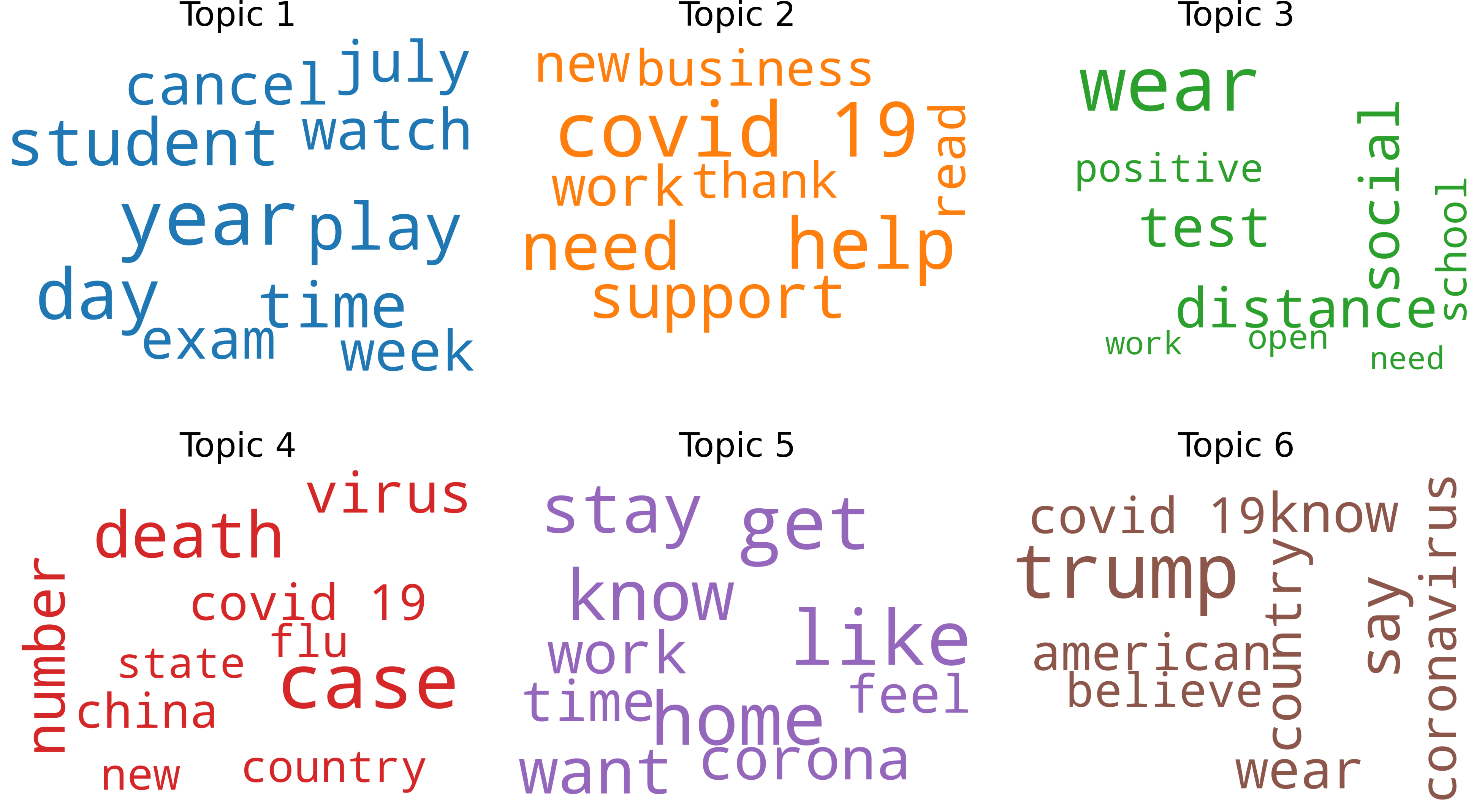}
  \caption{June 16 - July 15, 2020}
  \label{fig:sdJuneJuly}
\end{subfigure}
\begin{subfigure}{.33\textwidth}
  \centering
  \includegraphics[width=\linewidth]{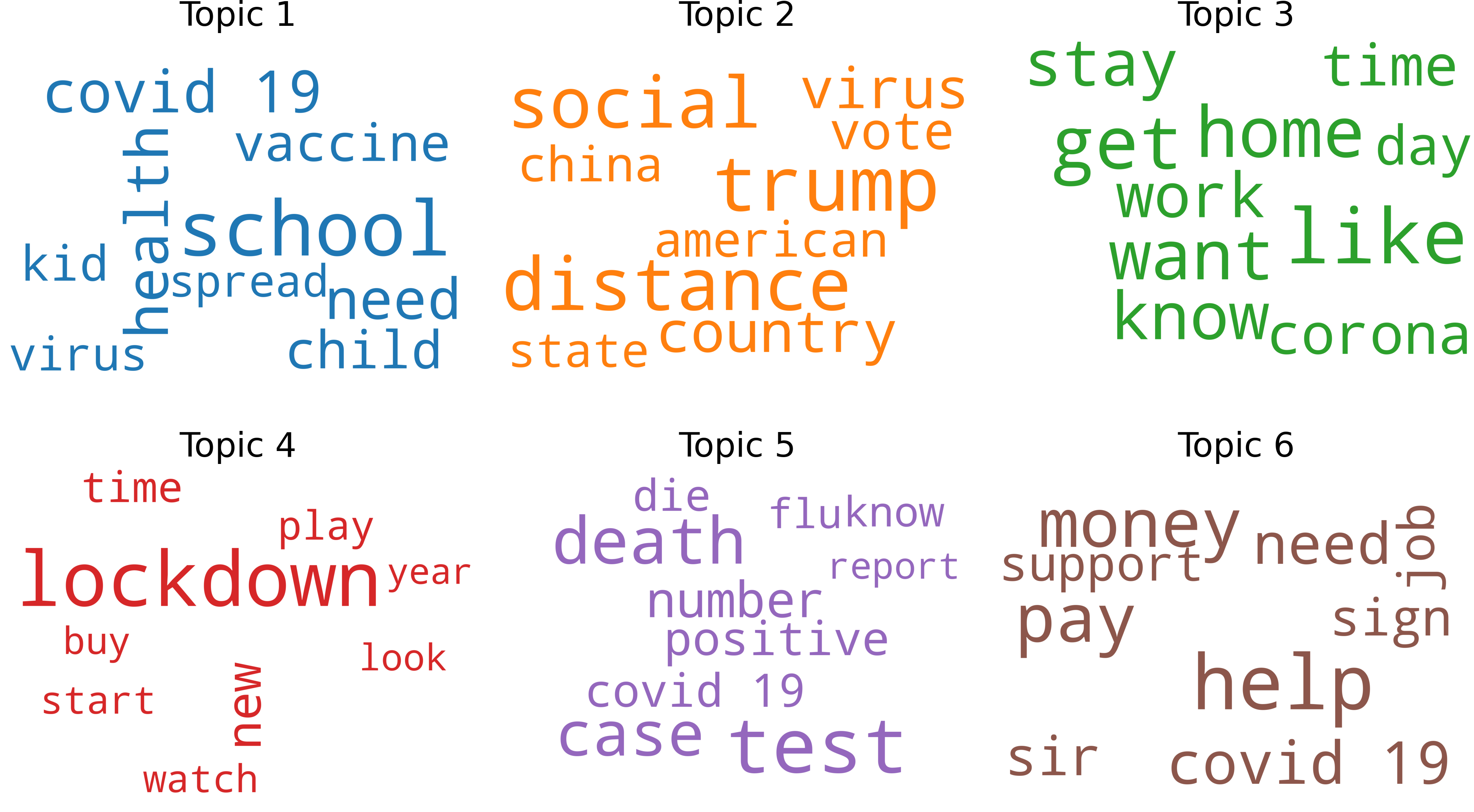}
  \caption{July 16 - August 15, 2020}
  \label{fig:sdJulyAugust}
\end{subfigure}
\caption{Topical Comparison of Self-Disclosing Coronavirus Tweets: May 16 - August 15, 2020 }
\label{COVIDSD_May_August}
\end{figure*}

\begin{figure*}[tbh]
\centering
\begin{subfigure}{.5\textwidth}
  \centering
  \includegraphics[width=.85\linewidth]{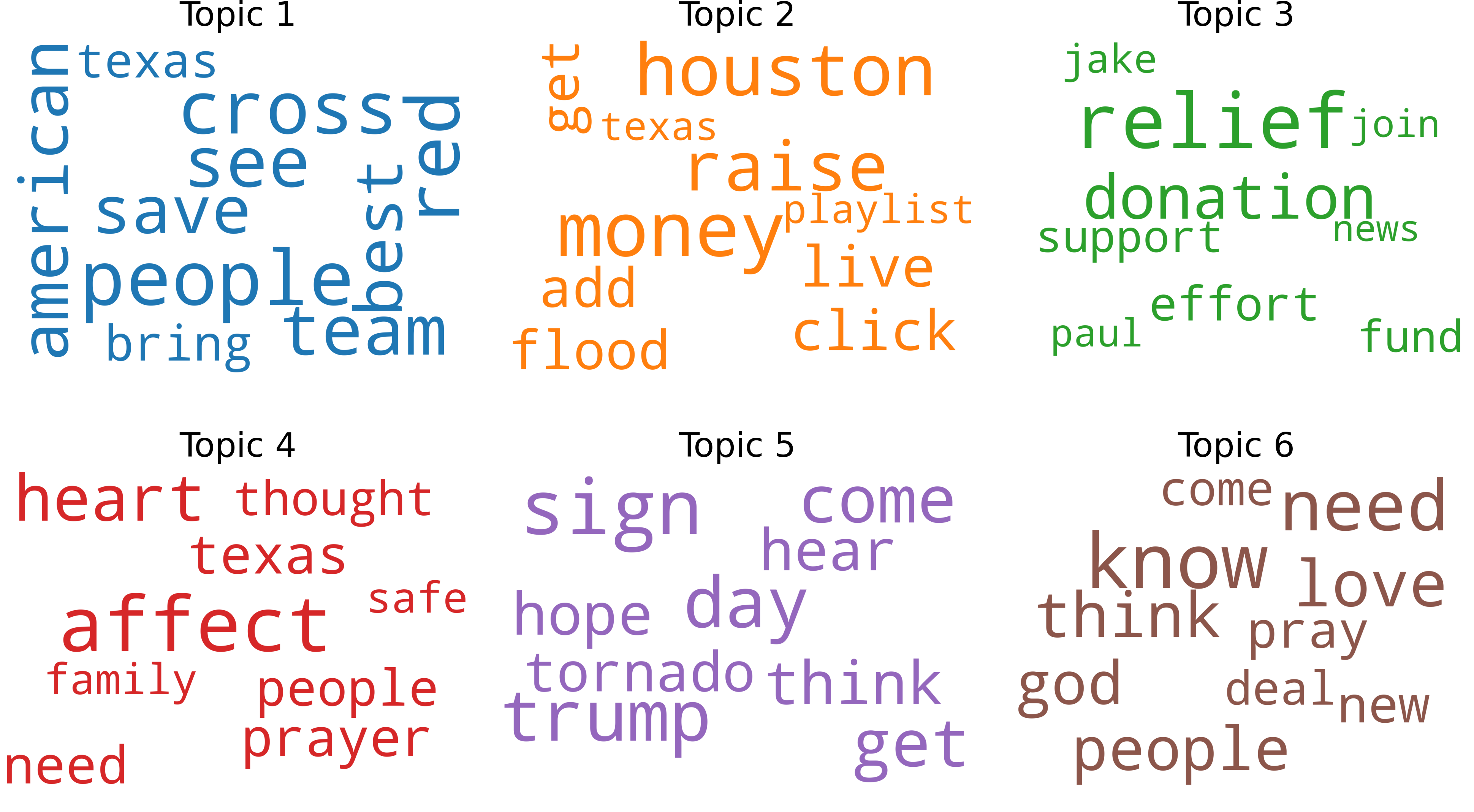}
  \caption{Self-disclosing Tweets}
  \label{fig:sub1_harvey}
\end{subfigure}%
\begin{subfigure}{.5\textwidth}
  \centering
  \includegraphics[width=.85\linewidth]{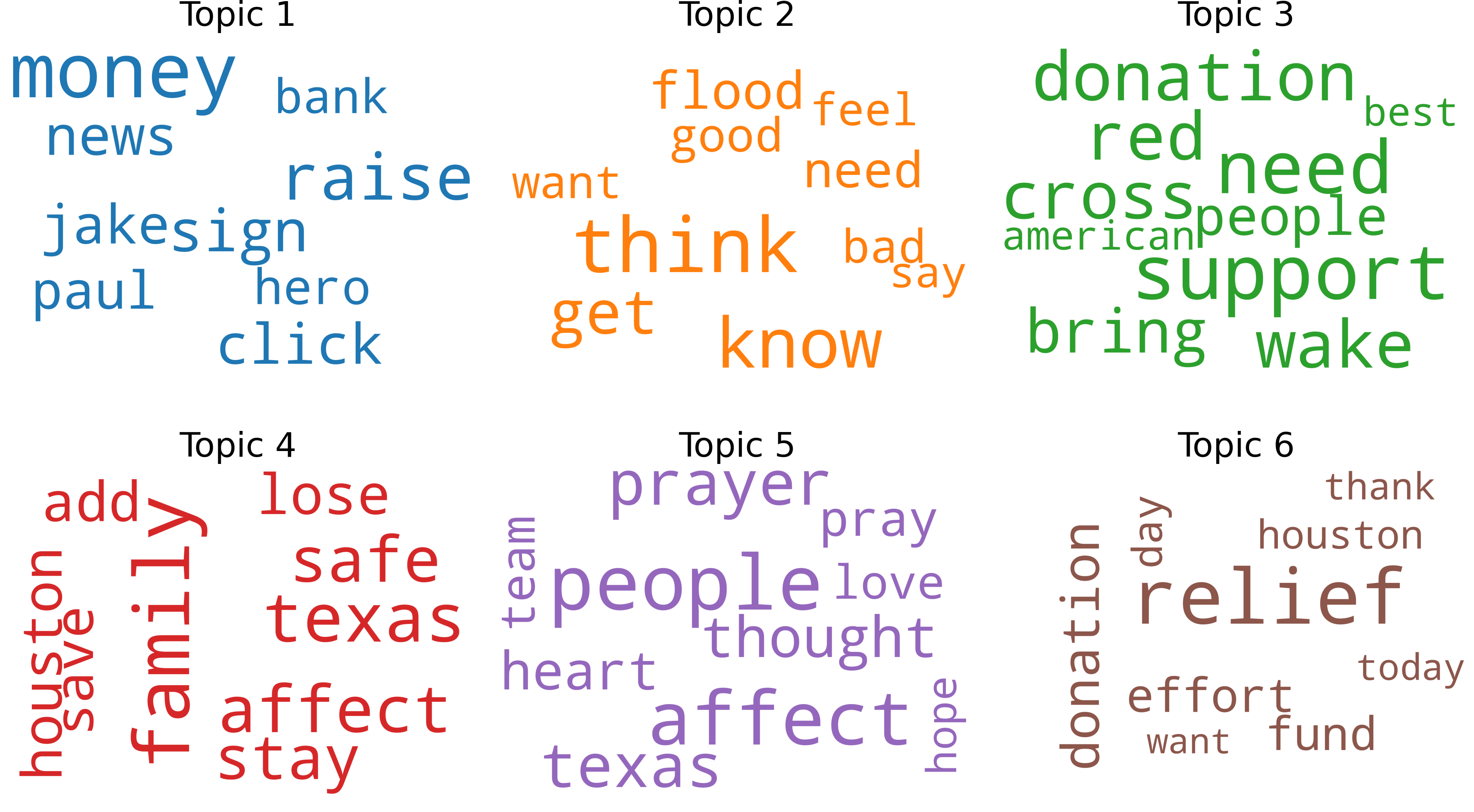}
  \caption{Non-self-disclosing Tweets}
  \label{fig:sub2_harvey}
\end{subfigure}
\caption{Topical Comparison of Hurricane Harvey Tweets}
\label{HarveySD}
\end{figure*}

We further examined messaging around the the Coronavirus pandemic through topic modeling, as outlined in Section \ref{topics}. As discussed, self-disclosing behaviors increased steeply around March 11th, and interesting distinctions are noticeable in the topical breakdown comparing the periods just before (January 21 - March 11) and after (March 12 - May 15) this date of interest, as reported in Figure \ref{COVIDSD_prePost}.  
Leading up to March 11th (Figure \ref{sd_PreMar11}), self-disclosing conversation focused on general information (Topic 1) and sentimental impacts (Topic 3) represented 20\% and 24\% of all Tweets, respectively. After March 11 (Figure \ref{sd_PostMar11}), terms related to global and sentimental aspects of the crisis remain present (Topic 3, Topic 2) but prominent conversations also shift to managing the spread of the virus (Topic 1, Topic 4) and discussions about needs, help, thanks and support (Topic 5, Topic 6). In fact, Topics 2 and 4 together make up the majority of the Tweets in that period, representing 23\% and 24\% of activity, respectively. This represents a shift from outward-looking to self-centric messaging as well as early evidence of emotional support and support seeking through disclosure.

Centered on the mid-March increase in disclosure behaviors, we also compare topical variance between disclosing and non-disclosing Coronavirus-related Tweets for entirety of the aforementioned subsetted time window (January 21 - May 15; see Figure \ref{COVIDSD_subSD_nonSD}). Generally, extracted topics reflect terminology pervasive in mainstream media at the onset of the crisis, including but not limited to expected impacts of COVID-19 (health, market, lockdown, quarantine), recommendations (stay home, mask, cancel, school), and political impact (Trump, administration, democrat). There is no noticeable distinction between topics extracted from the subset of Tweets containing self-disclosure and those without.

We also analyzed monthly subsets within the remaining timeline of our dataset, May 15 through August 15 (Figure \ref{COVIDSD_May_August}), where the proportion of self-disclosing conversations remains elevated, and in fact, continues to gradually increase higher over time. The observed sustained higher rate of self-disclosure through the summer is particularly interesting. We speculate that users may have recalibrated their own sharing practices during the pandemic, and that this recalibration may represent a longer-term effect.  Consistent with preceding time windows were themes of support seeking and political opinion. We observe the introduction of new themes related to social movements and education, while, discussions surrounding daily life (working/staying home) become the most prominent topic representing 38\% and 41\% of all self-disclosing Tweets in the final two months of the collection. This again suggests a possible transition from discussions of anxiety and need to potentially coping with lifestyle changes and establishing a new normal. 


\section{Comparative analysis}

For context, we compared observed self-disclosure during the Coronavirus pandemic to observed self-disclosure during Hurricane Harvey (2017). Although hurricanes are an annual expectation, the landfall duration and subsequent impact of Hurricane Harvey created a crisis throughout communities in the South Central region of the United States. This overwhelmed traditional emergency response infrastructure and affected citizens took to social media to seek emergency assistance \cite{sebastian2017hurricane,smith2018social}.

We consider a collection of 6,732,546 Tweet IDs representing posted content inclusive of keywords “Hurricane Harvey”, “Harvey”, and/or “HurricaneHarvey” during the 12-day period August 25, when Harvey first made landfall, through September 5, 2017 \cite{firoj2018twitter}. Similar to the Coronavirus dataset, we passed the set of Hurricane Harvey-related Tweet IDs through the Hydrator Tweet Retrieval Tool (v2.0)\footnote{https://github.com/DocNow/hydrator}, a sister desktop application to Twarc. We experienced a 33\% loss during data reconstitution (compare to 9.21\% loss in the Coronavirus-related dataset), attributable to Tweet and account deletion during the nearly 3 years which have passed. We filtered the resulting 4,379,462 Tweets for original content in the same fashion as we handled the Coronavirus data -- removing all quoted Tweets, retweets, content from verified accounts, and non-English content as identified by Twitter. The resulting 551,061 Tweets were passed through pre-processing and unsupervised labeling to detect instances of self-disclosure, and through topic analysis as detailed in Sections \ref{detection} and \ref{topics}. 

Across all 551,061 Tweets in this subset, we observed an average 9\% self-disclosure rate (49,595 Tweets) over the 12-day collection period -- substantially lower than the 19.07\% observed in the Coronavirus-related dataset. Several factors might account for the difference. The current pandemic has been marked by efforts to maintain social distance and, we have proposed, increased levels of disclosure may be related to relationship-building with online cohorts. But what we may also be seeing is a reflection of a general trend toward greater self-disclosure in online social media over the course of nearly three years separating the two events.

With respect to topical focus, we see important parallels between the two crises. As illustrated in Figure \ref{fig:sub1_harvey}, self-disclosing Tweets revealed emotional messaging centered on seeking immediate spiritual, physical, and monetary support (Topics 1, 2, 3, 4, 6) with top terms including ``red cross'', ``raise money'', ``donation'', ``relief'', and ``prayer''. While present, these themes are less prominent in the non-self-disclosing data. This finding across both datasets suggests that support-seeking during crisis might be a driver of self-disclosure and play a meaningful role in users' sharing practices.

Also mirroring the Coronavirus dataset, we observe one self-disclosing topic representing politically-motivated conversation (Topic 5). 
While there is topical overlap between the two classes of Harvey-related Tweets, non-disclosing Tweets (Figure \ref{fig:sub2_harvey}) presented a focus on contextual information related to the crisis with relevant keywords ``flood'', ``Texas'', and ``Houston'' (Topic 4).

\section{Discussion}

Perhaps the most striking observation in the analyses we have described is early evidence of heightened and sustained levels of self-disclosure during the ongoing Coronavirus pandemic, as compared to observed disclosure during Hurricane Harvey. This global crisis is unprecedented in a number of ways, one being the scale and scope of human interaction through social media. Concerns about privacy have been at the center of discussion in popular press (see, e.g, \cite{ScienceMag,BBC,WallStJ}), but most of this conversation has been about privacy tradeoffs related to cellphone tracking and similar approaches to location surveillance and individual health monitoring in service to public health. 
Many are willing to sacrifice some privacy in hopes of stemming the spread of the disease and helping to accelerate the return to normalcy; others are not. Scholarly work has begun to propose ``privacy first'' decentralized approaches for COVID-related tracking and notification (see, e.g., \cite{PACT,Covid-Watch,DP3T}).

We aim, with this work, to engage the research community in the work of better understanding more subtle, voluntary self-privacy violations emergent in the Coronavirus pandemic and in crisis more generally. We know that, during Hurricane Harvey, individuals took to social media in search of immediate aid. Today, individuals are leaning on their online social communities for ongoing engagement and support. Isolation, economic uncertainty, and health-related anxiety pose serious threat to mental health and well-being \cite{holmes2020multidisciplinary,AcadMedSciences}, but the potential manifestations of psychological impact in the domain of voluntary self-disclosure are unknown. The existing literature hints at the role of mood and emotion in the privacy calculus, but these relationships have not be well-established. Our analyses suggest that the current pandemic and its effects may impact self-disclosure behaviors. An open question is whether these heightened oversharing practices will become a new norm, or whether they fade over time and we will ultimately return to pre-COVID baselines. If heightened self-disclosure sustains beyond this crisis, we might look to social theory for explanations. Threshold models of collective behavior (see, e.g., \cite{granovetter1978threshold,centola2018experimental,wiedermann2020network}) may offer insights. 
As the crisis continues to unfold in the next several months, we suggest that additional work should aim to further develop, refine and test these hypotheses.

\bibliography{Bibliography.bib}

\end{document}